# Electrical Control of Silicon Photonic Crystal Cavity by Graphene

*Arka Majumdar* [1,2][‡]*, Jonghwan Kim* [1,][‡]*, Jelena Vuckovic* [2]*, Feng Wang* [1,3,][*]

[1] Department of Physics, University of California at Berkeley, Berkeley, CA 94720, USA

[2] E.L. Ginzton Laboratory, Stanford University, Stanford, CA-94305, USA

[3] Materials Science Division, Lawrence Berkeley National Laboratory, Berkeley, CA 94720, USA

ABSTRACT :

Efficient conversion of electrical signal to optical signal in nano-photonics enables solid state integration of electronics and photonics. Combination of graphene with photonic crystals is promising for electro-optic modulation. In this paper, we demonstrate that by electrostatic gating a single layer of graphene on top of a photonic crystal cavity, the cavity resonance can be changed significantly. A ~2nm change in the cavity resonance linewidth and almost 400% (6 dB) change in resonance reflectivity is observed. In addition, our analysis shows that a graphene-photonic crystal device can potentially be useful for a high speed, and low power absorptive and refractive modulator, while maintaining a small physical footprint.

TEXT:

**Introduction: :** Integrating photonics and electronics on the same platform holds great promise for the future of high performance computing[1]. However one of the major challenges to bring optics and electronics together is the efficient conversion between the optical and electronic signal, i.e., designing fast, low power electro-optic modulators as well as fast, sensitive photo-detectors. Developing compact and power-efficient electro-optic modulator requires progress in two directions: new optical material with strong and field-tunable optical transitions, and new nanophotonic design with resonantly enhanced light-matter interactions in a small mode volume.

Optical properties of Graphene, a two-dimensional material with linear dispersion and zero electron density of states at the Dirac point, can be easily controlled by applying an electric field[2,3]. Hence graphene is a very attractive candidate for performing electro-optic modulation. Recent studies have demonstrated efficient electro-optic modulation in silicon waveguide integrated with graphene (switching contrast of 0.1 dB $\mu m^{-1}$)[4]. However, the size required for the waveguide to achieve a good switching contrast is large, as light needs to travel a long distance overlapping with graphene. One can greatly enhance the light-graphene interaction and reduce the size of the optical modulator by coupling graphene to a photonic crystal cavity with high quality (Q) factor. Indeed the excellent electrical transport properties and nanofabrication capability can enable such integrated device for ultrahigh speed and low power operation in a small footprint. Very recently it has been shown that graphene can modify the photonic crystal cavity resonance and give rise to interesting non-linear optical phenomena. However, no electrical control of such graphene-cavity system has been demonstrated[5,6]. Several other proposals involve using graphene in distributed Bragg reflector cavity for enhancing light-matter interaction[7], but the cavities used are of high mode volume, and lower quality factor (~20). In this paper, we report electrical control of a silicon photonic crystal cavity (quality factor of

~1000-1500 and mode volume ~$(\lambda/n)^3$ , n being the refractive index of the cavity material) through electrostatic gating of a monolayer graphene on top of it. We show that although graphene is only one atom thick, its effect on the photonic crystal cavity is remarkably strong: both the cavity resonance linewidth as well as the cavity reflection can be modulated significantly.

**Effect of graphene:** The experiments are performed with linear three hole defect (L3) silicon photonic crystal cavities fabricated in Silicon-on-Insulator platform. The device thickness d is 250nm, with photonic crystal lattice periodicity a = 450 nm and radius r = 90 nm. The two holes at the end of the cavities are shifted by 0.15a [8]. The photonic crystals are fabricated by electron-beam lithography, followed by plasma etching and finally removing the silicon oxide underneath to make a free standing silicon photonic crystal membrane. On top of the cavities we transferred a large-area graphene grown by chemical vapor deposition using the standard growth and transfer processes[9,10]. For electrostatic gating of graphene we used a top electrolyte gating with ion-gel[11]. The device schematic is shown in Figure 1a. Figures 1b and 1c are scanning electron micrograph images of a fabricated photonic crystal cavity before and after the graphene transfer, respectively.

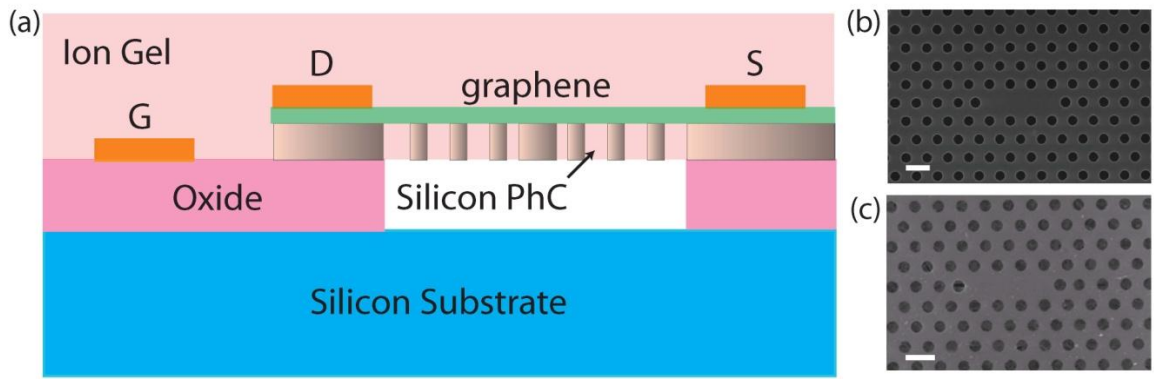

*Figure 1: (a) The schematic of the photonic crystal cavity-graphene device. The cavity is fabricated on a silicon-on-insulator platform, and the electrical gating is performed by means of contacts (Drain, Source and Gate) covered by an ion-gel. (b),(c): Scanning electron micrograph (SEM) of the fabricated cavities (b) without and (c) with graphene. The scale bars correspond to a 500nm distance.*

We characterize the photonic crystal cavities in cross-polarized reflectivity measurement setup with a broadband light-source (a super-continuum laser), where the cavity is kept at a 45$^o$ angle, the incident probe laser is vertically polarized, and we collect horizontally polarized light[12]. Collected light is analyzed by a spectrometer equipped with an InGaAs array detector. A quality (Q) factor of ~1000-1500 is observed for the fabricated cavities without graphene. We note that although a much higher quality factor can be obtained in a silicon photonic crystal cavity, we want to keep the quality factor moderate to achieve a relatively large spectral bandwidth. With graphene on top, the Q-factor drastically reduces to ~300-500, as shown in Figure 2a. The significant broadening of the cavity linewidth arises from graphene absorption. We measured the reflectivity spectrum from several cavities with slightly different r/a ratio. For all the cavities we consistently observe linewidth broadening (Figure 2b). The quantitative values of the broadening varies slightly in the range of 2 nm to 4 nm, consistent with our theoretical estimation (supplementary information) and the recent experimental observations[6]. Change in the cavity resonance frequency is quite small, and no consistent behavior is observed (Figure 2c).

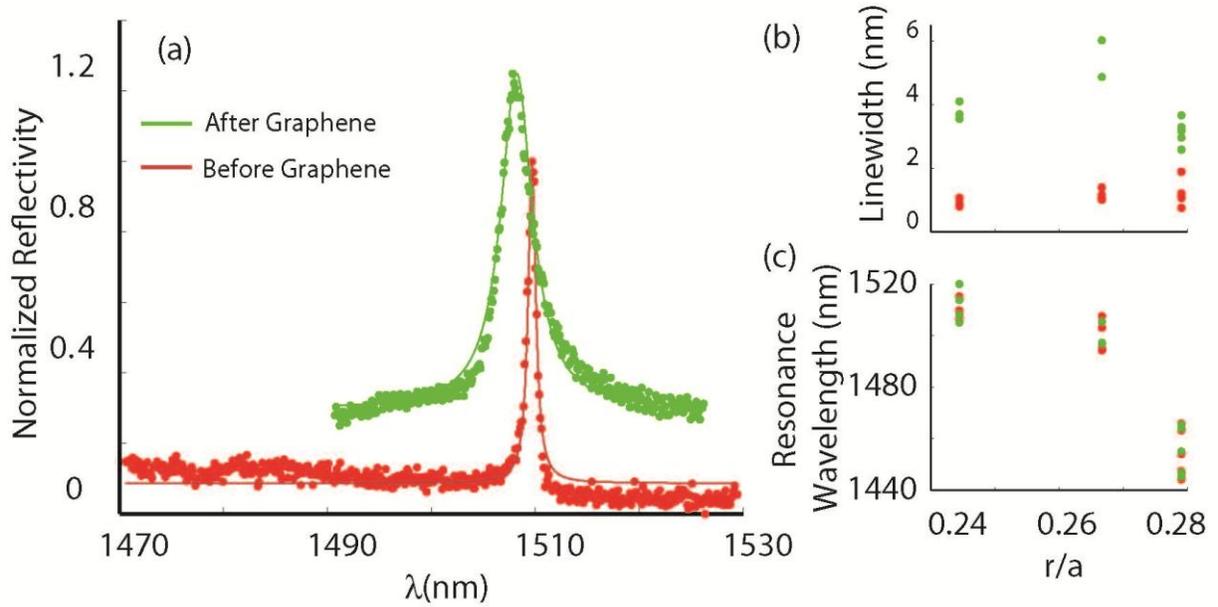

*Figure 2: Cavity reflectivity spectrum measured in the cross-polarized reflectivity setup: (a) Normalized reflectivity spectrum from a cavity before (red plot) and after (green plot) graphene transfer. Significant broadening of the cavity is observed due to graphene absorption. The two plots are vertically offset for clarity. (b) The linewidths of cavities as a function of r/a of the photonic crystal measured for several different cavities in three sets of cavities with different r/a. The uncertainty in the linewidth from the fitting is less than 0.2nm. (c) The cavity resonance wavelength as a function of r/a of the photonic crystal. The green and red dots, respectively, correspond to the situations after and before graphene transfer (both in in (b) and (c)).*

**Effect of electric field**: Then we study the effect of the electric field on the graphene-cavity device. We gate the graphene layer by means of the ion-gel, that we spin-coat on the device. Figure 3 shows the effect of the electric field on the graphene-cavity device. We simultaneously measure the cavity reflection spectrum and graphene resistance while varying the gate voltage at a step of 10 mV/sec. Figure 3a shows the cavity reflectivity spectra for different voltages. A narrowing in the cavity linewidth as well as an increase in the cavity reflectivity is clearly observed with increasing gating of the graphene. We fit all the spectra with a Lorentzian line-shape to extract the cavity resonance frequencies and cavity linewidths. Figure 3b shows the peak value of the cavity reflectivity as a function of the applied voltage. Figures 3c, d show the cavity linewidths and the resonance frequencies as a function of the applied voltage. Figure 3e shows the resistance between the drain and the source as a function of the gate voltage. From the transport data (Figure 3e) we clearly observe that graphene charge neutral point (the point where the resistance is maximum) is at around 0.5 V. The small deviation from the 0V is due to slight p-doping of graphene during the graphene transfer process. We observe narrowing of the cavity linewidth and an increase in the cavity reflection consistent with the fact that increased gating of the graphene reduces its absorption. The change in the cavity resonance is relatively small. We confirmed that in this voltage range, the ion gel does not affect the cavity resonance (data not shown here).

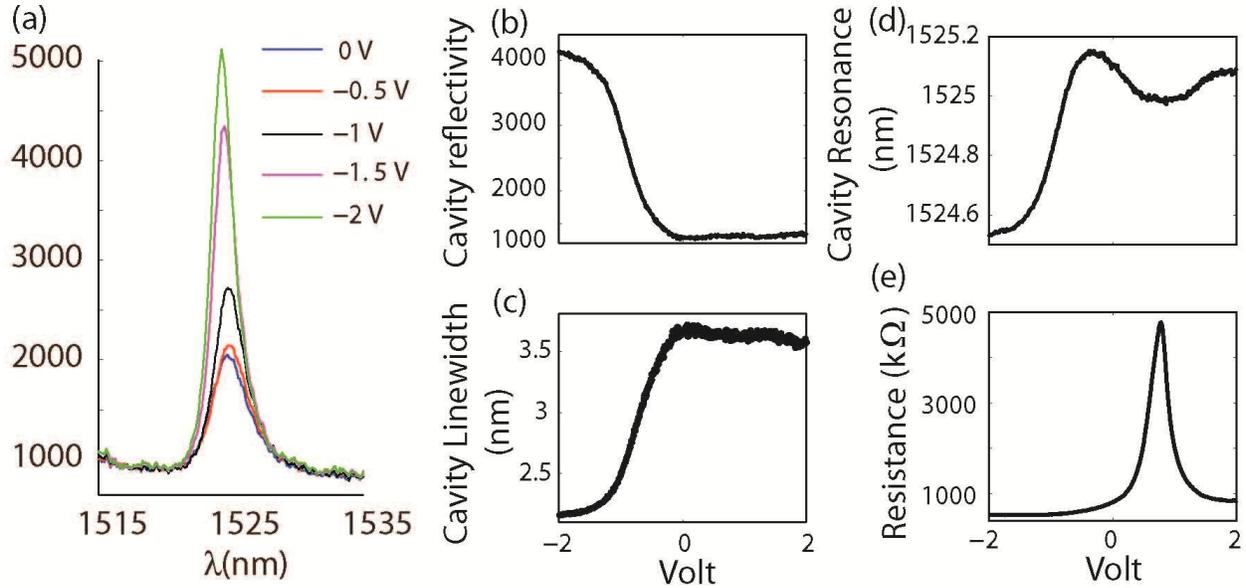

*Figure 3: Effect of electric field on graphene-cavity device: (a) Cavity reflectivity spectra for several voltages. An increase in cavity reflectivity and narrowing of cavity linewidth is observed with increased gating of the graphene. (b) The peak reflectivity of the cavity; (c) the cavity*

*linewidth; (d) the cavity resonance and (e) the resistance measured between the source and the drain as a function of the gate voltage.*

**Theory**: To conclusively prove that the change in cavity resonances are solely due to graphene, we theoretically fit our experimental observations. Gate-dependent complex dielectric constant of graphene has been extensively studied previously[13-15]. The complex dielectric function $\varepsilon_g(\omega)$ can be obtained from the optical conductivity $\sigma(\omega) = \sigma_1(\omega) + i\sigma_2(\omega)$ of graphene by using: $\varepsilon_g(\omega) = 1 + i\sigma(\omega)/(\omega\varepsilon_o d_g)$, where $d_g$ is the thickness of the graphene layer (we used $d_g \sim 1$nm in our fit). Under random phase approximation and using Kramer-Kronig relation, we can write the real and imaginary part of the optical conductivity as:

$$\sigma_1(\omega) = f(\omega) + \frac{q^2 \Gamma}{2\pi\hbar^2} \frac{1/\tau}{\omega^2 + (1/\tau)^2} \log\left(2\cosh\frac{2E_F}{\Gamma}\right)$$

$$\sigma_2(\omega) = -\frac{q^2}{8\hbar} \frac{2\omega}{\pi} \int_0^\infty \frac{f(\omega') - f(\omega)}{\omega'^2 - \omega^2} d\omega + \frac{q^2 \Gamma}{2\pi\hbar^2} \frac{\omega}{\omega^2 + (1/\tau)^2} \log\left(2\cosh\frac{2E_F}{\Gamma}\right)$$

with

$$f(\omega) = \frac{q^2}{8\hbar}\left[\tanh\left(\frac{\hbar\omega + 2|E_F|}{\Gamma}\right) + \tanh\left(\frac{\hbar\omega - 2|E_F|}{\Gamma}\right)\right]$$

and q is the electronic charge and $\Gamma$ is the interband transition broadening (estimated to be 150 meV from the fitting). The free carrier scattering rate $1/\tau$ can be neglected because it has little effect on the dielectric constants at the vicinity of the cavity resonance energy ($E_r = \hbar\omega_r$). From these equations we find that $\sigma_1(\omega)$ decreases when $2|E_F|$ is larger than $E_r$ and blocks the relevant interband transitions. $\sigma_2(\omega)$ has significant contribution both from intraband and interband transition. Contribution from the intraband transition $\left(\frac{q^2\Gamma}{2\pi\hbar^2}\frac{\omega}{\omega^2+\left(\frac{1}{\tau}\right)^2}log\left(2cosh\frac{2E_F}{\Gamma}\right)\right)$ increases monotonically with increasing $|E_F|$, i.e., with increasing carrier density. On the other hand, interband transition contribution has a minimum at $2|E_F| = E_r$. We note that in our experiment we are applying a voltage V to the graphene layer, and the Fermi level of graphene and applied voltage can be related by the formula[16,17]

$$E_F = \hbar v_f \sqrt{\pi(n_o + \frac{C|V|}{q})}$$

where, C is the effective capacitance per unit area, $v_f$ is the Fermi velocity for graphene and $n_o$ is the intrinsic carrier concentration. The effective capacitance of ion gel gating is C ~ 20 mF/m$^2$ [11,13,18,19]; Fermi velocity $v_f = 10^6$ m/s and neglected any intrinsic carrier concentrations. Figures 4a,b show the cavity linewidth and the cavity resonance wavelength as a function of the applied voltage. We found that the shift (both blue and red) and the line-width increase of the cavity resonance scale linearly with the gate-dependent dielectric constant of graphene. The theoretical fits in Figure 4a,b (red solid line) are obtained by the following relations: for the cavity linewidth $\Gamma_R = \Gamma_R^0 + \alpha Im[\varepsilon_g(\omega)]$ and for the cavity resonance wavelength $\lambda_R = \lambda_R^0 + \beta Re[\varepsilon_g(\omega)]$. Our simple model reproduces nicely the significant gate-induced decrease of cavity line-width, as well as the shift (both blue and red) in the cavity resonance frequencies. We note that the linewidth narrowing observed by gating graphene is ~1.5nm, which matches the theoretical estimate of the change in the linewidth just due to the graphene absorption (supplementary materials).

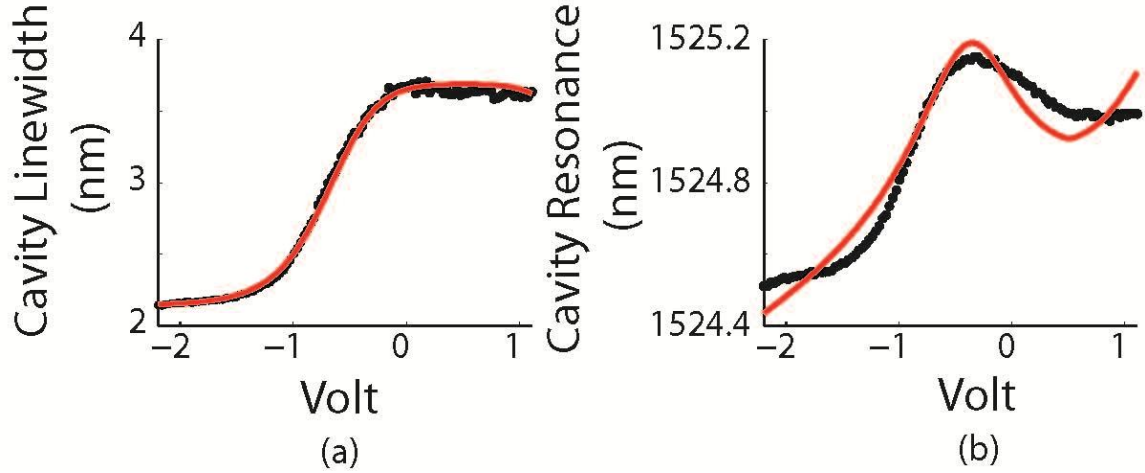

*Figure 4: Theoretical fit to the experimental data: (a) cavity linewidth and (b) cavity resonance as a function of the applied voltage. The decreasing cavity linewidth can be explained by the imaginary part of the graphene dielectric constant; whereas the change in the cavity resonance can be explained by the real part of the graphene dielectric constant.*

**Performance of the modulator**: In this paper, we have used ion-gel to electro-statically dope graphene to demonstrate the capability to modulate the cavity resonance by gating graphene. However, ion-gel gating cannot be used in high speed electro-optic modulator due to the slow response of ions under the electric field. Instead, one should employ semiconductor field-effect transistor structure to gate graphene and achieve ultrafast electro-optic modulation. Here, we theoretically analyze the performance (speed and energy consumption) of such a graphene-photonic crystal cavity modulator with silicon gating. Assuming we need a specific carrier density of $n_c$ (charge/area) to cause significant change in the graphene absorption, we can write $n_c = CV/q$, where, q is the electronic charge, V is the applied voltage and $C = \varepsilon/d$ is the capacitance per unit area, and d is the oxide thickness. The total capacitance will be $C_T = CA$, with A being the gated graphene area. Hence the total energy $E_T$ consumption for the modulator will be $1/2\, C_T V^2$. This means

$$E_T = \frac{1}{2}CAV^2 = \frac{Aq^2n_c^2d}{2\epsilon}$$

In a graphene-PhC modulator we can achieve: A = 1 µm$^2$ (assuming the graphene covers only the cavity region), d=5 nm, dielectric constant of Alumina is 10, and the carrier density required to do such modulation is $10^{17}$ /m$^2$, then the energy of the modulator can be around 8 fJ. The speed of the device will be limited by the RC constant of the device. The resistance of the device will come mostly from the graphene resistance, assuming the silicon is highly doped, and is of very low resistance. The dc conductivity of graphene in strongly doped region, as used in the modulator condition, can be quite high. Sukang Bae et. al. has shown a sheet resistivity of 30Ω/sq[20]. Assuming the length and width of the graphene electrode are ~1 micron, we find that the resistance of the device can be as low as 30Ω and the capacitance of the device is $C_T$ =17 fF. This makes the speed of the modulator to be ~250 GHz.

**Conclusion**: In summary, we observe drastic change in the parameters of a photonic crystal cavity by electrostatic gating of the graphene on top of it. A linewidth narrowing of 1.5 nm is observed, along with around 1nm shift in cavity resonance and a 6dB change in cavity reflectivity. Our analysis shows that such graphene-cavity based modulator can be used for very low power (fJ) electro-optic modulation maintaining a speed of 100's of GHz. We believe that such electrically controlled graphene-cavity device will enable technologies that benefit from local fast electrical tuning of cavities.

ASSOCIATED CONTENT

**Supporting Information**.

AUTHOR INFORMATION

**Corresponding Author**

* To whom correspondence should be addressed. Email:fengwang76@berkeley.edu

**Author Contributions**

A.M. and F.W. designed the experiment; A.M. and J.K. carried out the optical measurements, sample growth, fabrication and characterization. A.M, J.K., J.V. and F.W. performed theoretical analysis. A.M wrote the paper with inputs from all the authors. ‡These authors contributed equally.

**Funding Sources**

Optical characterization of this work was mainly supported by Office of Basic Energy Science, Department of Energy under Contract No. DE-AC02-05CH11231; photonic crystal fabrication and characterization are supported by the Office of Naval Research (PECASE Award; No: N00014-08-1-0561); and graphene synthesis and graphene-photonic crystal integration is supported by ONR MURI award N00014-09-1-1066. This work was performed in part at the Stanford Nanofabrication facility supported by the National Science Foundation.

ACKNOWLEDGMENT

We thank R. Segalman and B. Boudouris for providing the ion-gel; Sufei Shi, Long Ju, Yaqing Bie and Will Regan for help in preparing the sample and Gary Shambat for helpful discussion.

# Supplementary Material:

*Theoretical estimation of broadening of cavity linewidth due to graphene absorption:* The total optical energy $E_T$ stored in a cavity is given by $E_T = \varepsilon_o \varepsilon_r |E|^2 V_m$, where E is the electric field of the light, $V_m$ is the mode volume of the cavity, $\varepsilon_o$ and $\varepsilon_r$ are respectively, the vacuum permittivity and the relative permittivity of the material inside the cavity. The electric field inside the cavity (with resonance frequency $\omega_o$ and quality factor Q) decreases with time as $E(t)=E(0)\exp(-\omega_o t/2Q)$. If the bare cavity has a linewidth of $\Delta\omega_o = \omega_o/Q$ and the cavity with graphene on top has a linewidth of $\Delta\omega$, then the energy lost due to graphene over an infinitesimally small time-duration T is given by $\varepsilon_o \varepsilon_r |E(0)|^2 V_m (\exp(-\Delta\omega_o T) - \exp(-\Delta\omega T))$. This corresponds to the energy that is absorbed by graphene. The optical power absorbed in graphene (with conductivity $\sigma$) is given by $\sigma |E|^2 A_m$ where, E is the electric field sensed by graphene, and $A_m$ is the area where cavity field overlaps with graphene. Assuming graphene covers the whole cavity, we can approximate $A_m = V_m / d$, d being the thickness of the photonic crystal membrane. As the graphene is on top of the cavity, it senses the cavity field that evanescently coupled out of the top surface. Assuming an exponential fall off of the electric field along the cavity membrane from the center of the cavity the electric field sensed by graphene will be $E=E(0)\exp(-nd/2\lambda)$, where $\lambda$ is the cavity resonance wavelength, and n is the refractive index of the cavity material. Hence equating the two expressions of total energy lost at an infinitesimally small time duration T, we can write

$$\varepsilon_o \varepsilon_r |E(0)|^2 V_m \left(\exp(-\Delta\omega_o T) - \exp(-\Delta\omega T)\right) = \frac{\sigma |E(0)|^2 V_m T}{d} \exp\left(-\frac{nd}{\lambda}\right)$$

Expanding the exponential to the first order in T, we find that

$$\Delta\omega - \Delta\omega_o = \frac{\sigma}{d\varepsilon_o\,\varepsilon_r}\exp\left(-\frac{nd}{\lambda}\right)$$

with $\sigma = q^2/4\hbar$ for graphene, where q is the charge of an electron. In our device, d=250nm, the material used is silicon ($\varepsilon_r = 11.68$), and the center wavelength is ~1500nm. Hence theoretically estimated change in the cavity linewidth just due to the graphene absorption is ~1.6 nm.

***Effect of ion-gel***: The cavity resonance red shifts by 40-50 nm due to ion-gel, but the resonance linewidth does not change significantly (the quality factor of the cavity with ion-gel is slightly higher than without ion-gel). Figure S1 shows the normalized cavity reflectivity with and without ion-gel. However, no change in cavity resonance is observed as a function of the electric field.

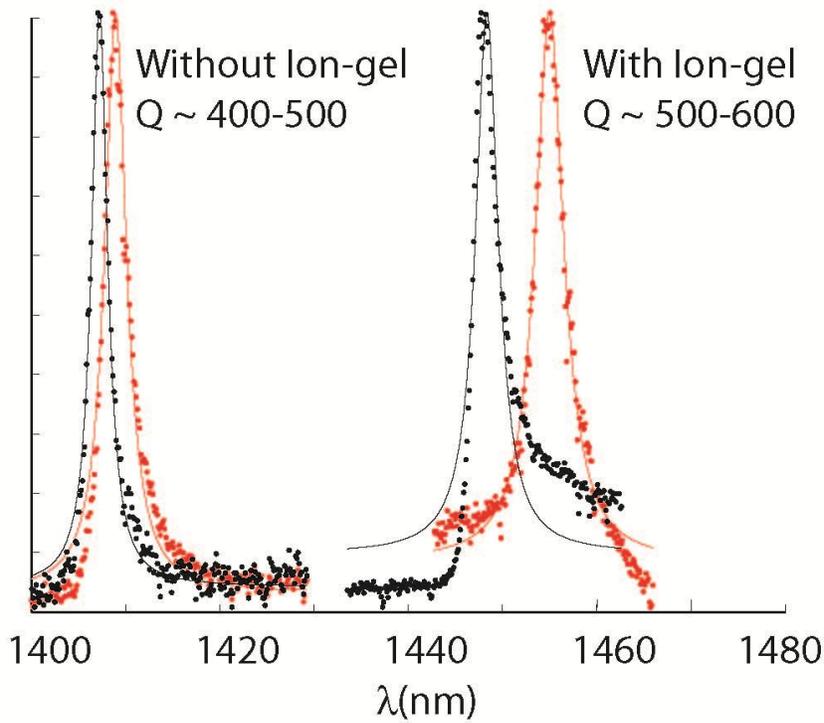

*Figure S1: The normalized cavity reflectivity with and without ion-gel.*